\documentclass[a4paper,12pt]{article}
\usepackage[utf8]{inputenc}
\usepackage{amssymb}
\usepackage{graphicx,amssymb,amsfonts}
\newlength{\extraspace}
\newlength{\extraspaces}
\RequirePackage[T1]{fontenc}
\RequirePackage{color,soul}
\RequirePackage{amssymb}
\RequirePackage[figtopcap]{subfigure}
\RequirePackage{array,multirow,graphicx}
\newcommand{\be}{\begin{equation}
\addtolength{\abovedisplayskip}{\extraspaces}
\addtolength{\belowdisplayskip}{\extraspaces}
\addtolength{\abovedisplayshortskip}{\extraspace}
\addtolength{\belowdisplayshortskip}{\extraspace}}
\newcommand{\ee}{\end{equation}}

\newcommand{\ba}{\begin{eqnarray}
\addtolength{\abovedisplayskip}{\extraspaces}
\addtolength{\belowdisplayskip}{\extraspaces}
\addtolength{\abovedisplayshortskip}{\extraspace}
\addtolength{\belowdisplayshortskip}{\extraspace}}
\newcommand{\ea}{\end{eqnarray}}

\newcommand{\nonu}{\nonumber \\[.5mm]}
\newcommand{\A}{&\!\!\!}

\begin{document}

\begin{center}
{\bf Analytic rotating black hole solutions in $N$-dimensional $f(T)$ gravity }
\end{center}
\centerline{\bf G.G.L. Nashed$^{1,2,3}$ and W. El Hanafy$^{1,3}$\footnote{nashed@bue.edu.eg\\ \hspace*{.7cm}waleed.elhanafy@bue.edu.eg}}

\centerline{{$^{1}$Centre for theoretical physics, the British University in Egypt, 11837 - P.O. Box 43, Egypt}}
\centerline{{$^{2}$Mathematics Department, Faculty of Science, Ain Shams University, Cairo, Egypt}}
\centerline{{$^{3}$Egyptian Relativity Group (ERG)}}

\hspace{2cm} \hspace{2cm}
\\
\\
\\
\\
A non-diagonal vielbein ansatz is applied to the $N$-dimension field equations of $f(T)$ gravity. An analytical vacuum solution is derived for the quadratic polynomial $f(T)=T+\epsilon T^2$ and an inverse relation between the coupling constant $\epsilon$ and the cosmological constant $\Lambda$. Since the induced metric has off-diagonal components, that cannot be removed by a mere of a coordinate transformation, the solution has a rotating parameter. The curvature and torsion scalars invariants are calculated to study the singularities and horizons of the solution. In contrast to the general relativity (GR), the Cauchy horizon  differs from the horizon which shows the effect of the higher order torsion. The general expression of the energy-momentum vector of $f(T)$ gravity is used to calculate the energy of the system. Finally, we have shown that this kind of solution satisfies the first law of thermodynamics in the framework of $f(T)$ gravitational theories.
\section{Introduction}\label{S1}
General relativity is constructed in the Riemann's geometry that reduces Minkowski spacetime in the absence of gravitational field. In the Riemannian geometry, the differentiable manifold is described by a metric tensor which measures the distances between different space points. However, gravitation in this theory is a manifestation of the curvature of the spacetime, where Riemann Christoffel tensor is responsible for this curvature. All quantities in this manifold can be defined in terms of the metric tensor \cite{Ea}. At present time, the matter to extend GR to include torsion is considered as an urgent issue because many questions  depend on whether the spacetime connection is symmetric or not. General relativity is an orthodox  theory that does not allow quantum effects.  However, these effects should be taken into account in any theory that accompanies with gravity. Changing geometry from $V_4$ (4-dimensional Riemannian manifolds) to $U_4$ (4-dimensional Weitzenb\"ock manifolds)  is considered as a first direct  extension that attempts  to incorporate the spin fields of matter into the same geometrical scheme of GR. As an example of this, is  the mass-energy in which curvature is the source  of gravity  while the spin is  of torsion.  Einstein-Cartan-Sciama-Kibble theory is considered as  one of the most significant trails in this direction . Nevertheless, the task of spin-matter fields does not explain the function of torsion tensor that seems to possess an important tasks  in any fundamental theory.

The accelerated expansion of our Universe is confirmed by many separated cosmological experiments \cite{RFC}--\cite{Fa}. This expansion can be explained in GR by taking into account the dark energy component in the total energy of our Universe. It is widely accepted, without real justifications, to interpret these components by inserting a cosmological constant $\Lambda$ into Einstein's field equation which is called $\Lambda$CDM model \cite{CL}. Also, this model can be extended by adding a speculated dynamical fluid with a negative pressure, e.g. quintessence fluid. Therefore, we can consider the accelerated expansion of our Universe as an indicator of the failure of our information to the principles of gravitational field and thus a modification of GR is necessary. This modification can be done by making the action as a function of the curvature scalar $R$, i.e., $f(R)$ \cite{CF}--\cite{SF}.

Using a different processor we can setup another method and use the Weitzenb\"ock connection which includes torsion  as an alternative of curvature. This process was suggested by Einstein who called it ``teleparallel equivalent of general relativity'' (TEGR) \cite{Ea,HS9}--\cite{Mj}. This theory, TEGR, is closely related to GR and differs only by a total derivative  term  in the action. The dynamical objects in such geometry are the four linearly independent vierbeins\footnote{In this study Greek indices and Latin ones refer to  the coordinate and tangent spaces respectively.} $b^{a}{_{\mu}}$. The advantage of this framework is that the torsion tensor is formed solely from the products of first derivatives of the vielbein. In TEGR formalism, the features of gravitational field are described in torsion scalar \cite{HS9, M1}. Therefore, guided by the construction of $f(R)$ gravitational theory we are allowed to postulate
the Lagrangian of the amended Einstein-Hilbert action as a function of the torsion scalar to extend TEGR in a natural way. This extension is called $f(T)$, where $T$ is the teleparallelism scalar torsion, gravitational theory \cite{FF}--\cite{Wh}. It is necessary to note that in $f(T)$ theory the field equations are of second-order while the field equations of $f(R)$ are of fourth-order. Different aspects of cosmology in $f(T)$ theory  have been discussed in the literature \cite{LSB}--\cite{GW}.

A black hole solution \cite{CGSV} has been investigated in the quadratic polynomial $f(T)$ gravity using a diagonal vielbein as a first approach to study the feature of the theory. As is well known, the diagonal relation between vielbein and the metric is always allowed in the linear teleparallel gravity. However, the non-diagonal relation in the generalized theories, e.g. $f(T)$ gravity is an essential. The aim of the present study to derive a rotating $N$-dimensional black hole solution for a quadratic polynomial $f(T)$ gravitational theory using a non-diagonal vielbein and discuss its relevant physics. This study is arranged  as follow: In Section \ref{S2}, installation of teleparallel space and the $f(T)$ gravitational theory are provided. In Section \ref{S3}, a vielbein field\footnote{In 4-dimensional spacetime case, we call the vielbein a vierbein or tetrad.} is applied to $N$-dimension field equations of $f(T)=T+\epsilon T^{2}$ and exact rotating solution is delivered. This solution is asymptotically de Sitter or anti de Sitter (dS/AdS). The main merit of this solution is that its scalar torsion is constant. In  Section \ref{S5}, some relevant physics, singularities, energy and first law of thermodynamics, are discussed. Final section is reserved to the discussion.
\section{Installing $f(T)$  gravitational theories}\label{S2}
We are going, in this section, to briefly review $f(T)$ gravitational theory. Then we apply the field equations of a quadratic polynomial $f(T)=T+\epsilon T^2$ to a non-diagonal vielbein field with a spherical symmetry aiming to find a new black hole solution.
\subsection{Teleparallel space}\label{Sec2.1}
In this subsection, we give a brief survey of the teleparallel space. In the literature, this space has many names, e.g. distant parallelism, Weitzenb\"{o}ck, absolute parallelism (AP), vielbein, parallelizable etc. For more details, we refer reader to \cite{M62}--\cite{M2013}. Teleparallel space is a pair $(M,\,b_{a})$, where $M$ is an $N$-dimensional manifold and $b_{a}$ ($a=1,\cdots, N$) are $N$ independent vector fields defined globally on $M$. The vector fields $b_{a}$ are known as the parallelization vector fields or vielbein fields.

Let $b_{a}{^{\mu}}$ $(\mu = 1, ..., N)$ be the coordinate components of the $a$-th vector field $b_{a}$. Einstein summation convention are applied on both Greek (world) and Latin (mesh) indices. The covariant components $b^{a}{_{\mu}}$ of  the vector $b_{a}$  and its contravariant  ones  satisfy the following  orthogonal condition
\begin{equation}\label{orthonormality}
b_{a}{^{\mu}}b^{a}{_{\nu}}=\delta^{\mu}_{\nu}\quad \textmd{and}\quad b_{a}{^{\mu}}b^{b}{_{\mu}}=\delta^{b}_{a},
\end{equation}
with  $\delta$ being the Kronecker tensor. Due to the linear independence of the vector $b_{a}$, the determinant $b:=\det (b_{a}{^{\mu}})$ must be  nonzero.

On a teleparallel spacetime, $(M,\,b_{a})$, uniquely, there exists a non symmetric linear (Weitzenb\"{o}ck) connection, which is constructed from the vielbein fields as
\begin{equation}\label{W_connection}
\Gamma^{\alpha}{_{\mu\nu}}:=b^{a}{_{\mu}}\partial_{\nu}b_{a}{^{\alpha}}=-b_{a}{^{\alpha}}\partial_{\nu}b^{a}{_{\mu}}.
\end{equation}
This connection identifies the important property
\begin{equation}\label{AP_condition}
\nabla^{(\Gamma)}_{\nu}b_{a}{^{\mu}}:=\partial_{\nu}
{b_a}^\mu+{\Gamma^\mu}_{\lambda \nu} {b_a}^\lambda\equiv 0,
\end{equation}
which is known as the AP-condition. So the connection (\ref{W_connection}) sometimes can be called the canonical connection. Associated to it, the covariant differential operator $\nabla^{(\Gamma)}_{\nu}$ is given as (\ref{AP_condition}).

The curvature tensor $R^{\alpha}{_{\epsilon\mu\nu}}$ and the torsion tensor $T^{\epsilon}{_{\nu\mu}}$ of the canonical connection can be calculated using the non commutation relation of an arbitrary vector fields $V_{a}$ is given by
$$\nabla^{(\Gamma)}_{\nu}\nabla^{(\Gamma)}_{\mu}V_{a}{^{\alpha}} - \nabla^{(\Gamma)}_{\mu}\nabla^{(\Gamma)}_{\nu}V_{a}{^{\alpha}} = R^{\alpha}{_{\epsilon\mu\nu}}
V_{a}{^{\epsilon}} + T^{\epsilon}{_{\nu\mu}} \nabla^{(\Gamma)}_{\epsilon} V_{a}{^{\alpha}}.$$
Combining the above non commutation relation and the AP-condition (\ref{AP_condition}), the curvature tensor $R^{\alpha}_{~~\mu\nu\sigma}$ of the canonical connection $\Gamma^{\alpha}_{~\mu\nu}$ must be null identically \cite{M62}. Moreover, we can induce the metric tensor corresponding to a given vielbein fields as
\begin{equation}\label{metric}
g_{\mu \nu} := \eta_{ab}b^{a}{_{\mu}}b^{b}{_{\nu}}
\end{equation}
and the contravariant one
\begin{equation}\label{inverse}
g^{\mu \nu} = \eta^{ab}b_{a}{^{\mu}}b_{b}{^{\nu}}.
\end{equation}
Also, we can construct the symmetric linear (Levi-Civita) connection from the metric tensor $g_{\mu\nu}$ as
\begin{equation}\label{Christoffel}
 {\mathring \Gamma}{^{\alpha}}{_{\mu\nu}}= \frac{1}{2} g^{\alpha \sigma}\left(\partial_{\nu}g_{\mu \sigma}+\partial_{\mu}g_{\nu \sigma}-\partial_{\sigma}g_{\mu \nu}\right).
\end{equation}
Using Eqs. (\ref{AP_condition}) and (\ref{Christoffel}), it is easy to check that both, the canonical connection $\Gamma{^\alpha}{_{\mu\nu}}$ (\ref{W_connection}) and the Leivi-Civita connection, are metric ones, i.e.
$$\nabla^{(\Gamma)}_{\sigma}g_{\mu\nu}\equiv 0, \quad \nabla^{(\mathring{\Gamma})}_{\sigma}g_{\mu\nu}\equiv 0,$$
where the operator $\nabla^{(\mathring{\Gamma})}_{\sigma}$ is the covariant derivative associated with the Levi-Civita connection $\mathring{\Gamma}$. For the non symmetric connection (\ref{W_connection}), we define the torsion tensor
\begin{equation}
T^\alpha{_{\mu\nu}}:={\Gamma^\alpha}_{\nu\mu}-{\Gamma^\alpha}_{\mu\nu}={b_a}^\alpha\left(\partial_\mu{b^a}_\nu
-\partial_\nu{b^a}_\mu\right).\label{Torsion}\\
\end{equation}
Also, we define the contortion tensor $K^{\alpha}_{~\mu\nu}$ as the difference between Weitzenb\"{o}ck and Levi-Civita connections as
\begin{equation}
K^{\alpha}{_{\mu\nu}} := \Gamma^{\alpha}_{~\mu\nu} - \mathring{\Gamma}{^{\alpha}}_{\mu\nu}=b_{a}{^{\alpha}}~ \nabla_{\nu}b^{a}{_{\mu}}. \label{contortion}
\end{equation}
Due to the  symmetrization of the connection $\mathring{\Gamma}{^{\alpha}}{_{\mu\nu}}$, torsion and contortion tensors are related by
\begin{equation}\label{torsion-contortion}
T^{\alpha}{_{\mu\nu}} = K^{\alpha}{_{\mu\nu}} - K^{\alpha}{_{\nu\mu}}.
\end{equation}
Alternatively, one can write torsion in terms of contortion as
\begin{equation}\label{contortion03}
        K_{\alpha \mu \nu}=\frac{1}{2}\left(T_{\nu\alpha\mu}+T_{\alpha\mu\nu}-T_{\mu\alpha\nu}\right),
\end{equation}
where $T_{\sigma\mu\nu} = g_{\epsilon\sigma}\,T^{\epsilon}_{~\mu\nu}$\, and \,$K_{\mu\nu\sigma} =
g_{\epsilon\mu}\,K^{\epsilon}_{~\nu\sigma}$.  The torsion $T_{\sigma\mu\nu}$ (contortion $K_{\mu\nu\sigma}$) tensor is skew-symmetric in the last (first) pair of indices. Moreover, Eqs. (\ref{torsion-contortion}) and (\ref{contortion03}) show that the torsion tensor vanishes if and only if the contortion tensor vanishes.
\subsection{$f(T)$ gravitational theory}\label{S1.1}
In $f(T)$ gravity, and similarly to all torsional formulations, we use the vierbein fields $b^{i}{_{\mu}}$ which form an orthonormal base.  The Lagrangian of TEGR i.e.
the torsion scalar $T$, is constructed by contractions of the torsion tensor as  \cite{WN}
\begin{equation}\label{Tor_sc1}
T=\frac{1}{4} T^{\mu \nu \lambda}T_{\mu \nu \lambda}+\frac{1}{2} T^{\mu \nu \lambda}T_{\lambda \nu \mu }-
T^{\mu \nu}{}_ \mu T_{\lambda \nu}{}^ \lambda,
\end{equation}
Similar to  the $f(R)$ extensions of GR  we can extend $T$ to a function $f(T)$,
constructing the action of $f(T)$ gravity \cite{FF,BF9}:
\begin{equation}\label{q7}
{\cal L}=\frac{1}{2\kappa}\int |b|(f(T)-2\Lambda)~d^{N}x+\frac{1}{2\kappa}\int |b|{\cal L}_{Matter}~d^{N}x,
\end{equation}
where $\kappa$  is the N-dimensional gravitational constant given by $\kappa =2(N-3)\Omega_{N-1} G_N$, where  $G_N$  being  Newton’s constant in $N$–dimensions and
$\Omega_{N-1}$ is the volume of an $(N-1)$-dimensional unit sphere, which is given by
the expression $\Omega_{N-1} = \frac{2\pi^{(N-1)/2}}{\Gamma((N-1)/2)}$ (with the $\Gamma$-function
of the argument that depends on the dimension of spacetime)\footnote{When $N = 4$, one can show that $2(N-3)\Omega_{N-1} = 8 \pi$.}.
$ |b|=\sqrt{-g}=\det\left({b^a}_\mu\right)$ and $ {\cal L}_{Matter}$ is the Lagrangian matter. The variation of Eq. (\ref{q7}) with respect to the vielbein field ${b^i}_\mu$ and its first derivative  gives the following field equations \cite{BF9}
\begin{eqnarray}\label{q81}
{S_\mu}^{\rho \nu} \partial_{\rho} T f_{TT}+\left[b^{-1}{b^i}_\mu\partial_\rho\left(b{b_i}^\alpha
{S_\alpha}^{\rho \nu}\right)-{T^\alpha}_{\lambda \mu}{S_\alpha}^{\nu \lambda}\right]f_T
+\frac{f-2\Lambda}{4}\delta^\nu_\mu =\frac{1}{2} \kappa{{\cal
T}}^\nu{}_\mu,\end{eqnarray}
with $f := f(T)$, \ \   $f_{T}:=\frac{\partial f(T)}{\partial T}$, \ \  $f_{TT}:=\frac{\partial^2 f(T)}{\partial T^2}$ and ${\cal
T}^\nu{}_\mu$ is the
energy momentum tensor.

Equation  (\ref{q81}) can rewritten as
\be \partial_\nu \Biggl[b{S}^{a \rho \nu} f(T)_T\Biggr]=\kappa b
{b^a}_\mu \Biggl[t^{\rho \mu}+{{\cal
T}}^{\rho \mu}\Biggr],\ee

where $t^{\nu \mu}$  is defined as \be t^{\nu
\mu}:=\frac{1}{\kappa}\Biggl[4f(T)_T {S^\alpha}^{\nu
\lambda}{T_{\alpha \lambda}}^{\mu}-g^{\nu \mu} f(T)\Biggr].\ee
 Due to the anti- symmetrization of the tensor  ${S}^{a \nu \lambda}$ we get \be\label{q9}
\partial_\mu \partial_\nu\left[b{S}^{a \mu \nu} f(T)_T\right]=0, \quad
 \textrm{that \quad gives}
\quad \partial_\mu\left[b\left(t^{a \mu}+{{\cal
T}}^{a \mu}\right)\right]=0. \ee
Equation (\ref{q9}) yields the continuity equation in the form \be \label{qq1}
\frac{d}{dt}\int_V d^{(N-1)}x \ b \ {b^a}_\mu \left(t^{0 \mu}+{\cal
T}^{0 \mu}\right) + \oint_\Sigma \left[b \ {b^a}_\mu \ \left(t^{j
\mu}+{\cal T}^{j\mu}\right)\right]d\Sigma_{j}=0,\ee
where the integration is on ($N-1$) volume $V$ bounded by the surface $\Sigma$. This recommends the quantity $t^{\lambda \mu}$ to represent the energy-momentum tensor of the gravitational field in
the frame of $f(T)$ gravitational theories \cite{US3}. Therefore,
the total energy-momentum tensor of  $f(T)$ gravitational theory  is defined as \be P^a:=\int_{V} d^{(N-1)}x\;
\ b \ {b^a}_\mu \left(t^{0 \mu}+{{\cal
T}}^{0 \mu}\right)=\frac{1}{\kappa}\int_{V} d^{(N-1)}x \; \partial_\nu\left[b{S}^{a 0
\nu} f(T)_T\right].\ee Equation (18) is  the generalization of the energy-momentum tensor that can be used to calculate energy and spatial momentum. It is clear that the TEGR case \cite{MDTC} is recovered by setting $f(T)=T$.
\section{Rotating  solution in $f(T)$ gravitational theories}\label{S3}
For an $N$-dimensional spacetime having spherical symmetry with a flat horizon, we write the following vielbein ansatz in cylindrical coordinates ($t$, $r$, $\phi_1$, $\phi_2$ $\cdots$ $\phi_{N-3}$, $z$) as \cite{CGSV}
\begin{eqnarray}\label{tetrad}
\nonumber \left({b^{i}}_{\mu}\right)=\left(
  \begin{array}{cccccc}
    S_1(r) & 0 & 0 & 0&\cdots & S_2(r)  \\[5pt]
    0&S_3(r) &0 &0 &\cdots &0 \\[5pt]
    0&0 & \frac{r}{\lambda} & 0 &\cdots &0  \\[5pt]
      0&0 & 0&\frac{r}{\lambda}  &\cdots &0  \\[5pt]
        \vdots & \vdots & \vdots &\vdots&\ \vdots \cdots &\vdots  \\[5pt]
 c_2r &  0  &0&0  &\cdots &c_1 r\\
  \end{array}
\right).&\\
\end{eqnarray}
where $\lambda\equiv \sqrt{\frac{-(N-1)(N-2)}{2\Lambda}}$, $S_i(r), \; i=1\cdots 3$ are three unknown functions of the radial coordinate $r$, while $c_1$ and $c_2$ are some constants\footnote{The vielbein (\ref{tetrad}) is not the most general non-diagonal one. However, we use it to facilitate the calculations.}. It is clear that, if the two constants $c_1=1$, $c_2=0$ and the function $S_{2}(r)=0$  the vielbein (\ref{tetrad}) reduces to the previous  diagonal vielbein case studied in \cite{CGSV}. The line element of (\ref{tetrad}) takes the form
\be
ds^2=(S_1{}^2-c_2{}^2 r^2)dt^2-S_3{}^2dr^2-\left(\frac{r^2}{\lambda^2}\right)\sum_{i=1}^{i=N-3} d\phi_i{}^2 -(c_1{}^2 r^2-S_2{}^2)dz^2-2(c_1c_2r^2-S_1S_2)dzdt\label{rotat_metric}.\ee  Equation  (\ref{rotat_metric}) arises from the vielbein ansatz (\ref{tetrad}) using  Eq. (\ref{metric}).
Substituting  (\ref{tetrad}) into (\ref{Tor_sc1}) we evaluate the torsion scalar in the form\footnote{For abbreviation we will write $S_i(r)\equiv S_i$,  $S'_i\equiv\frac{dS_i}{dr}$ .}
\begin{eqnarray}\label{df}
&& T=\frac{1}{2r^2S_3{}^2 (c_1S_1-c_2S_2)^2}\left(S_2{}^2S'_1{}^2-S'_1\Biggl[2S_1S_2S'_2+4(N-2)c_1 r\{c_1S_1-c_2S_2\}\right]+S_1{}^2S'_2{}^2\nonu
& &-4(N-2)rc_2S'_2(c_1S_1-c_2S_2)+2(N-2)(N-3)(c_1S_1-c_2S_2)^2\Biggr),
\end{eqnarray}
where $S'_i=\displaystyle\frac{d S_i(r)}{d r}$. Inserting the vielbein (\ref{tetrad}) into the field equations (\ref{q81}), and assuming the vacuum  case; we get the following non-vanishing components:
\begin{eqnarray}\label{df1}
&  & t\; t\;-comp.=\nonumber\\
&&\frac{1}{4r^3S_3{}^3(c_1S_1-c_2S_2)^3}\Biggl[f_{TT}T'S_3r(c_2S_2-c_1S_1)\Biggl(\{S_2[S_1{}^2-2r^2c_2{}^2]+r^2c_1c_2S_1\}S'_2+S_2(c_1c_2r^2-S_1S_2)S'_1\nonu
& &+2(2N-5)rc_1c_2S_1S_2-2(N-3)rc_2{}^2S_2{}^2-2(N-2)rc_1{}^2S_1{}^2\Biggr)-f_{T}\Biggl\{rS_3(c_1S_1-c_2S_2)\Biggl(\Biggl[S_2(S_1{}^2-2r^2c_2{}^2)\nonu
& &+r^2c_1c_2S_1\Biggr]S''_2 +S_2[r^2c_1c_2-S_2S_1]S''_1\Biggr)+rS_1S_3c_1(S_1{}^2-r^2c_2{}^2)S'_2{}^2+S'_2\Biggl[ rS_3(c_1S_1+c_2S_2)
\nonu
& &\times (r^2c_1c_2-S_2S_1)S'_1- (c_1S_1-c_2S_2)\Biggl\{r[S_2(S_1{}^2-2r^2c_2{}^2)+r^2c_1c_2S_1]S'_3- S_3\Biggl[(3N-8)r^2c_2c_1S_1\nonu
& &+(N-4)S_2S_1{}^2-2(2N-5)r^2c_2{}^2S_2\Biggr]\Biggr\}\Biggr]S'_2+rc_2S_2S_3(S_2{}^2-r^2c_1{}^2)S'_1{}^2-
S'_1\Biggl\{rS_2 S'_3 (r^2c_1c_2-S_1S_2)\nonumber\\
&&+S_3[2(N-2)r^2c_1{}^2S_1-3(N-2)r^2c_1c_2S_2+(N-4)S_1S_2{}^2]\Biggr\}(c_1S_1-c_2S_2)+2 (c_1S_1-c_2S_2)^2\Biggl([N-2]c_1S_1\nonumber\\
&&-[N-3]c_2S_2\Biggr)[rS'_3-(N-3)S_3]\Biggr\}-\frac{f-2\Lambda}{4}\; ,\nonu
&  & t\; z\;-comp.=\nonumber\\
& & \frac{1}{4r^3S_3{}^3(c_1S_1-c_2S_2)^3}\Biggl[f_{TT}rT'S_3(c_1S_1-c_2S_2)\Biggl([2r^2c_1c_2S_1-S_2S_1{}^2-r^2c_2{}^2S_2]S'_1-S_1\Biggl[S'_2(c_2{}^2r^2-S_1{}^2)\nonu
& &+2rc_2(c_1S_1-c_2S_2)\Biggr]\Biggr)+f_{T}\Biggl[S_3r(c_1S_1-c_2S_2)\{[2r^2c_1c_2S_1-S_2S_1{}^2-r^2c_2{}^2S_2]S''_1-S_1(c_2{}^2r^2-S_1{}^2)S''_2\}\nonu
& &+rS_2S_3(2c_2S_1S_2-c_1S_1{}^2-r^2c_1c_2{}^2)S'_1{}^2+S'_1\Biggl[rS_3\left(r^2c_2{}^3S_2+c_1S_1{}^3
+r^2c_1c_2{}^2S_1-3c_2S_2S_1{}^2\right)S'_2
\nonu
& &-(c_1S_1-c_2S_2)\left\{r[2r^2c_1c_2S_1-S_2S_1{}^2-r^2c_2{}^2S_2]S'_3-[2(N-3)r^2c_1c_2S_1S_3-(N-4)\{r^2c_2{}^2S_2+S_2S_1{}^2\}]\right\}\Biggr]
\nonu
& &+S_1\Biggr\{rS'_2{}^2c_2(S_3S_1{}^2-r^2c_2{}^2S_3)+S'_2(c_1S_1-c_2S_2)[rS'_3(r^2c_2{}^2-S_1{}^2)-(N-2)r^2c_2{}^2S_3+(N-4)S_3S_1{}^2]\nonu
& &+2rc_2(c_1S_1-c_2S_2)^2[rS'_3-(N-3)S_3]\Biggr\}\Biggr],\nonu
&  & r\; r\;-comp.=\nonumber\\
& &\frac{f_T}{4r^2S_3{}^2(c_1S_1-c_2S_2)^2}\Biggl(S_2{}^2S'_1{}^2+2S'_1[2(N-2)rc_1(c_1S_1-c_2S_2)-S_1S_2S'_2]+S_1{}^2S'_2{}^2\nonumber\\
& & -4(N-2)rc_2S'_2(c_1S_1-c_2S_2)+2(N-2)(N-3)(c_1S_1-c_2S_2)^2\Biggr)-\frac{f-2\Lambda}{4},\nonumber\\
&&\nonu
&& \nonu
&&\nonu
&  & \phi_1\; \phi_1=\phi_2\; \phi_2=\phi_3\; \phi_3=\cdots =\phi_{N-3}\; \phi_{N-3}\;-comp.\nonumber\\
&&\frac{1}{2r^2S_3{}^3(c_1S_1-c_2S_2)}\Biggl[f_{TT}T'rS_3(rc_1S'_1-rc_2S'_2+(n-3)[c_1S_1-c_2S_2])+f_{T}\Biggl\{r^2c_1S_3S''_1-r^2c_2S_3S''_2\nonumber\\
&  &-rS'_3\Biggl(rc_1S'_1-rc_2S'_2+(N-3)[c_1S_1-c_2S_2]\Biggr)+S_3[(2N-5)\{rc_1S'_1-rc_2S'_2\}+(N-3)^2\{c_1S_1-c_2S_2\}]\Biggr\}\Biggr]\nonu
&&-\frac{f-2\Lambda}{4},\nonumber\\
\nonu
&  &z\; t\;-comp.=\nonumber\\
& & \frac{1}{4r^3S_3{}^3(c_1S_1-c_2S_2)^3}\Biggl[f_{TT}T'rS_3(c_2S_2-c_1S_1)\Biggl[S'_2\{S_1[r^2c_1{}^2+S_2{}^2]-2r^2c_1c_2S_2\}+S_2\Biggl([c_1{}^2r^2-S_2{}^2]S'_1\nonu
& &-2rc_1[c_1S_1-c_2S_2]\Biggr)\Biggr]-f_{T}\Biggl\{rS_3(c_1S_1-c_2S_2)\Biggl([S_1(r^2c_1{}^2+S_2{}^2)-2r^2c_1c_2S_2]S''_2+S_2S''_1[c_1{}^2r^2-S_2{}^2]\Biggr)\nonu
& &-rS_1S_3(r^2c_1{}^2c_2-2c_1S_1S_2+c_2S_2{}^2)S'_2{}^2+S'_2\Biggl\{r S_3(r^2c_1{}^3S_1+r^2c_1{}^2c_2S_2-3c_1S_1S_2{}^2+c_2S_2{}^3)S'_1 \nonu
& &-[c_1S_1-c_2S_2]\left(rS'_3[r^2c_1{}^2S_1-2r^2c_1c_2S_2+S_1S_2{}^2]-2(N-3)r^2c_1c_2S_2S_3+S_1S_3(N-4)[r^2c_1{}^2+S_2{}^2]\right)\Biggr\}\nonu
& &-S_2\Biggl(r c_1S_3(r^2c_1{}^2-S_2{}^2)S'_1{}^2+S'_1\Biggl[rS'_3(r^2c_1{}^2-S_2{}^2)-S_3[(N-2)r^2c_1{}^2-(N-4)S_2{}^2]\Biggr](c_1S_1-c_2S_2)\nonu
& &+2r c_1[rS'_3-(N-3)S_3](c_1S_1-c_2S_2)^2\Biggr)\Biggr\}\Biggr],\nonumber\\
&  &z\; z\;-comp.=\nonumber\\
& & \frac{1}{4r^3S_3{}^3(c_1S_1-c_2S_2)^3}\Biggl[f_{TT}rT'S_3(c_2S_2-c_1S_1)\Biggl[r^2c_1S'_1\{c_2S_2-2c_1S_1\}+r^2c_1c_2S_1S'_2-2(N-2)r c_2{}^2S_2{}^2\nonumber\\
&  & +2(2N-5)rc_1c_2S_1S_2-2(N-3)r c_1{}^2S_1{}^2+S_1S_2{}^2S'_1-S_2S_1{}^2S'_2\Biggr]+f_{T}\Biggl\{rS_3(c_1S_1-c_2S_2)\nonu
& &\times \Biggl(S''_1[(2r^2c_1{}^2-S_2{}^2)S_1-r^2c_1c_2S_2]-S''_2S_1[c_1c_2r^2-S_2S_1]\Biggr)-rS_2S_3c_2 (r^2c_1{}^2-S_2{}^2)S'_1{}^2\nonu
& &+S'_1\Biggl\{rS_3(c_1S_1+c_2S_2) (r^2c_1c_2-S_1S_2)S'_2-[c_1S_1-c_2S_2]\Biggl [rS'_3\{[2r^2c_1{}^2-S_2{}^2]S_1-r^2c_1c_2S_2\}\nonu
& &-r^2 c_1S_3\{2(2N-5)c_1S_1-(3N-8)c_2S_2\}-(N-4)S_1S_2{}^2\Biggr]\Biggr\}-rS_1S_3c_1 (r^2c_2{}^2-S_1{}^2)S'_2{}^2\nonu
& &+(c_1S_1-c_2S_2)[rS_1(r^2c_1c_2-S_1S_2)S'_3-\{r^2c_2S_3(N-2)[3c_1S_1-2c_2S_2]-(N-4)S_2S_1{}^2\}]S'_2\nonu
& &-2r[(N-3)c_1S_1-(N-2)c_2S_2](c_1S_1-c_2S_2)^2[rS'_3-(N-3)S_3]-\frac{f-2\Lambda}{4}\;.\end{eqnarray}
We constrain the system by the following
\be f(T)=T+\epsilon T^2, \qquad \qquad \textrm {and} \qquad \qquad \Lambda=\frac{1}{24\epsilon}. \ee
Here the second constraint is to simplify the results. However, the vacuum solution can be found in general. For more details about the second constraint, one may consult \cite{CGSV}. Substituting Eq. (23) into Eqs. (22), the black-hole solution is given as
\begin{eqnarray}\label{Solution}
& &  S_1(r)=c_1\sqrt{\frac{2(c_3-c_4r^{N-1})}{(N-1)r^{N-3}}}\;,\quad   S_2(r)=-c_2\sqrt{\frac{2(c_3-c_4r^{N-1})}{(N-1)r^{N-3}}}\;,\quad  S_3(r)=\sqrt{\frac{{6(N-1)(N-2)\epsilon c_4r^{N-3}}}{c_3-c_4r^{N-1}}}\;,\nonu
& &\end{eqnarray} where $c_3$ and $c_4$ are constants of integration. Equation (24) shows that when the constants $c_1=1$ and $c_2=0$, the solution reproduces the static black hole solution of \cite{CGSV}. We note that the null value of the constant $c_{2}$ implies the vanishing of the function $S_{2}(r)$. So we need not to impose the vanishing of $S_{2}(r)$ as an extra condition to reduce to the static black hole solution. To understand the characteristics of the above derived solution and its relevant physics, we are going to study the singularities, horizon, energy and first law of thermodynamics in the following section.
\section{Relevant Physics}\label{S5}
In this section, we are going to study some physical quantities of the analytical solution at hand.
\subsection{Singularities}\label{S1.1}
We start with studying one of the fundamental concepts in any gravitational theory, that is the singularity. Substituting solution (\ref{Solution}) into metric (\ref{rotat_metric}), the spacetime configuration takes the form
\ba \A \A ds^2=\frac{2c_1{}^2[c_3-c_4r^{N-1}]-c_2{}^2(N-1)r^{N-1}}{(N-1)r^{N-3}}dt^2-\frac{6(N-1)(N-2)\epsilon c_3 r^{N-3} }{c_3-c_4r^{N-1}}dr^2-\left(\frac{r^2}{\lambda^2}\right)\sum_{i=1}^{i=N-3} d\phi_i{}^2 \nonu
\A \A-\frac{2c_2{}^2[c_3-c_4r^{N-1}]-c_{1}^2(N-1)r^{N-1}}{(N-1)r^{N-3}}dz^2-\frac{c_2c_2\{2[c_3-c_4r^{N-1}]-(N-1)r^{N-1}\}}{(N-1)r^{N-3}}dtdz_i.\nonu
& & \ea
Equation (25) shows that the spacetime metric has a cross term that can not be removed by a coordinate transformation. This cross term is responsible for the rotation similar to the cross term appears in GR which creates Kerr black hole.  Equation (25) shows that when $c_1=1$ and $c_2=0$, we get
\ba \A \A ds^2=\frac{2[c_3-c_4r^{N-1}]}{(N-1)r^{N-3}}dt^2-\frac{6(N-1)(N-2)\epsilon c_3 r^{N-3} }{c_3-c_4r^{N-1}}dr^2-\left(\frac{r^2}{\lambda^2}\right)\sum_{i=1}^{i=N-3} d\phi_i{}^2 +\frac{c_1{}^2(N-1)r^{N-1}}{(N-1)r^{N-3}}dz^2,\nonu
& & \ea
which represents the gravitational field of a static black hole \cite{CGSV}.

To study spacetime singularities of the solution at hand, first we find the radial distance $r$ at which the functions $S_1$, $S_2$ and $S_3$ become null or infinitely at large. Since these functions are coordinate dependent quantities, we would like to make sure that the singular points are real and not a reflection of a bad choice of the coordinate system. In order to distinguish real singularities from coordinate ones, we study various invariants. These invariants do not change under coordinate transformation. If they are not defined at a specific spacetime point, they will be undefined at that point in any other coordinate choice. Then, the singular point will be physical singularity. In GR, we usually study some invariants, e.g. the Ricci scalar, the Kretschmann scalar, \textit{etc}, but all invariants are constructed from the Riemann tensor and its contractions. However, in TEGR, there are two ways to calculate invariants. In the first, one uses the obtained solution to calculate the torsion invariants, e.g. the torsion scalar $T$. In the second, one employs the solution to obtain the induced metric, then to calculate the curvature invariants. The comparison between these two ways is an interesting topic of this study which may shed out a light on the differences between the curvature based gravity and the torsion one. More specifically, one can use the vielbeins, Weitzenb\"ock's connection to construct all torsion based invariants, or to use the metric, Levi-Civita connection and construct all the curvature based invariants. Since particles follow geodesics defined by the Levi-Civita connection, some prefer to study the curvature invariants instead of the torsion ones. However, we can say that the two ways, (invariants constructed from metric and  those constructed from vielbeins), are equivalent only when a suitable choice of the vielbeins and the metric is selected. Therefore, classically, the above two ways might be alternatives, however, on the quantization level, it would be important to discover which is more fundamental field, the metric or the vielbein? We calculate some curvature and torsion invariants corresponding to the solution (\ref{Solution}) as given below,   using the symmetric  connection of Eq. (6) for curvature invariants and the non-symmetric one given by Eq. (2) for torsion invariants,  and get
\begin{eqnarray} \A \A R^{\mu \nu \lambda \rho}R_{\mu \nu \lambda \rho}=\frac{6Nc_4{}^2r^{2(N-1)}+3(N-2)^2(N-3)c_3{}^2}{108(N-1)(N-2)^2c_4{}^2\epsilon^2r^{2(N-1)}},\qquad  R^{\mu \nu}R_{\mu \nu} =\frac{N}{36(N-2)^2\epsilon^2},\qquad  R =-\frac{N}{6(N-2)\epsilon},\nonu
\A \A T^{\mu \nu \lambda}T_{\mu \nu \lambda} = \frac{4c_4{}^2r^{2(N-1)}-4c_3c_4r^{N-1}+(N-1)c_3{}^2}{12c_4\epsilon(N-2)r^{N-1}(c_4r^{N-1}-c_3)}, \qquad T^\mu T_\mu =\frac{(N-1)(2c_4r^{N-1}-c_3)^2}{24(N-2)r^{N-1}\epsilon c_4(c_4r^{N-1}-c_3)}, \nonu
\A \A T(r)=-\frac{1}{6\epsilon}\;,\qquad \qquad \nabla_\alpha T^\alpha=\frac{(N-1)}{6(N-2)\epsilon}\; \Rightarrow R=-T-2\nabla_\alpha T^\alpha.\nonu\end{eqnarray}
From the above calculations, we have the following comparison:
\begin{itemize}
\item [(i)] The invariants  $R^{\mu \nu \lambda \rho}R_{\mu \nu \lambda \rho}$, $ T^{\mu \nu \lambda}T_{\mu \nu \lambda}$ and $ T^{\mu}T_{\mu}$ have a  singularity at $r=0$.
\item [(ii)] The invariants  $ T^{\mu \nu \lambda}T_{\mu \nu \lambda}$ and $ T^{\mu}T_{\mu}$ have a more singularity at $c_3=c_4r^{N-1}$.
\item [(iii)] The scalars $R^{\mu \nu}R_{\mu \nu}$, $R$ and $T$  have no singularities and the black hole solution is regular even at $r=0$. We see that the limit $\epsilon=0$ is not valid to keep regular black hole solution.
\item [(iv)] All the above invariants are not defined when $\epsilon \rightarrow 0$. This means that solution (\ref{Solution}) has no analogy in GR and can't reduce to GR (or TEGR). Indeed, this result depends on the constraint  $\Lambda=\frac{1}{24 \epsilon}$. This constraint facilitate the calculations and make the above differential equations solvable.
\item [(v)] Equation (27) shows that Caushy horizon, $g_{tt}=0$ and the horizons constructed from $g^{rr}=0$ are the same which is familiar in GR (TEGR). However, this condition is not satisfied for Eq. (27). We may say that this is due to the contribution of the higher order torsion. This contribution appears in a non trivial black hole, like a rotating one. This point still needs more study which will be study elsewhere.
\end{itemize}
\subsection{Energy}\label{S1.1}
We next study the energy of the system according to the solution\footnote{In this study we take the Newtonian constant to be an effective constant in which $G_{eff.}=\frac{G_{N}}{f_T} $ \cite{WQM}.} (\ref{Solution}). Using Eq. (18)  we calculate the necessary components needed to study the evaluation of energy as:
\be S^{001}=\frac{c_2{}^2c_3-2^{(N-2)}c_4c_2{}^2r^{(N-1)}-6^{(N-2)/2}c_1{}^2 r^{(N-1)}}{(12)^{(N-2)}r^N\epsilon c_4(c_1{}^2-c_2{}^2)^2},  \ee
\ba P^0=E\A=\A \frac{2^{(N-2)}(N-2)\Omega_{N-1}\sqrt{2^{(N-2)/2}}(c_1{}^2-c_2{}^2)(c_3-c_4r^{(N-1)})}{\kappa(27)^{N-3}\sqrt{\epsilon c_4}}\nonu
\A=\A\frac{2^{(N-3)}(N-2)\sqrt{2^{(N-2)/2}}(c_1{}^2-c_2{}^2)(c_3-c_4r^{(N-1)})}{(N-3)G_N(27)^{N-3}\sqrt{\epsilon c_4}}, \ea
where the value of $\kappa$ has been used in the second equation of Eq. (29). The value of energy of the above equation is not a finite value therefore,  we must use the regularized method to get a finite value of the energy. This regularized expression takes the form
\be P^a:=\frac{1}{\kappa}\int_V d^{N-2}x  \left[b{S}^{a 0
0} f(T)_T\right]-\frac{1}{\kappa}\int_V d^{N-2}x \left[b{S}^{a 0
0} f(T)_T\right]_{physical\;  quantities\; equal\; zero}.\ee Using (30) in solution (24) we get
\ba E\A=\A\frac{2^{N-2}(N-2)\Omega_{N-1}\sqrt{2^{(N-2)/2}} (c_1{}^2-c_2{}^2)c_3}{\kappa(27)^{(N-2)/2}\sqrt{\epsilon c_4}}\nonu
\A =\A \frac{2^{N-3}(N-2)\sqrt{2^{(N-2)/2}} (c_1{}^2-c_2{}^2)c_3}{(N-3)G_N(27)^{(N-2)/2}\sqrt{\epsilon c_4}},\ea
which is a finite value.
\subsection{First Law of thermodynamics}\label{S1.1}
There is a great deal of work in analyzing the behavior of the horizon thermodynamics in modified  theories of GR. In a wide category of these theories, one gets solutions with horizons and can connect the temperature and entropy with horizons. Since temperature $T$ can be determined from periodicity of the Euclidean time,  determining the right form of entropy is the most non-trivial issue. Here we shall  briefly describe how these results arise in a class of theories which are natural modification of TEGR. A fundamental law need to be studied in modification of TEGR gravity is the fulfillment of the first law of thermodynamics. Miao et al. \cite{MLM} have discussed that if the first law of thermodynamics is satisfied within $f(T)$ gravitational theories or not. They split the non symmetric field equations (13) into symmetric and skew symmetric parts to have the forms
\begin{eqnarray}\label{q8}
\A \A L_{(\mu\nu)}:=S_{\mu \nu \rho} \partial^{\rho} T f_{TT}+f_T \left[G_{\mu \nu} -\frac{1}{2}g_{\mu \nu}T\right]
+\frac{f-2\Lambda}{2}g_{\nu \mu} =\kappa {\cal T}_{\nu \mu},\nonu
\A \A L_{[\mu \nu]}:=S_{[\mu \nu] \rho} \partial^{\rho} T f_{TT}=0.\end{eqnarray}
Assuming an exact Killing vector they have shown that for a heat flux $\delta Q$  passing through  black hole horizon and by using the symmetric part of Eq. (32) they show
\be  \delta Q=\frac{\kappa}{2\pi}\left[\frac{f_T dA}{4}\right]^{d\lambda}_0+\frac{1}{\kappa}\int_H k^\nu  f_{TT} \ T_{,\mu}(\xi^\rho S_{\rho \nu}{}^ \mu-\nabla_\nu \xi^\mu),\ee where $H$ stands for the black hole horizon which in this study is equal to $(N - 2)$-dimensional
boundary of the hypersurface at infinity.
It is shown that the first term in Eq. (33) can be rewritten as $T\delta S$ \cite{ MLM}. Therefore, when the second term in Eq. (33) is not vanishing there will be a violation of the first law of thermodynamics. Miao et al.  \cite{MLM} have shown that the second term can't be equal to zero. Therefore, if we need to satisfy the first law we must have either to have $f_{TT}=0$ which gives the TEGR (GR) theory, or to have $T=constant$. Indeed, solution (24) enforces the torsion scalar to be a constant. Therefore, the black hole solution (24) satisfies the first law of thermodynamics.
\section{Concluding remarks}
We show that finding an exact solution in  modified theories of gravity, e.g. $f(R)$ or $f(T)$, is not an easy task. In this work we derived an exact rotating black hole solution in the $f(T)$ gravity framework. Indeed, the authors of \cite{CGSV}, have found a static black hole solution using a diagonal vielbein. However, studying non-diagonal vielbein in $f(T)$ theories is necessary. In this work, we have employed a non-diagonal vielbein field, with three unknown functions, to the quadratic $f(T)=T+\epsilon T^2$ field equations to study the non-charged case. We obtained an exact solution by taking the the useful constraint $\Lambda=\frac{1}{24 \epsilon}$ into account.  The derived analytical solution containing two constants of integration. This solution coincides with what derived in \cite{CGSV} when the off-diagonal components are set equal to zero, i.e., $c_2=S_2=0$.

The solution of the non-diagonal vielbein is a new one and has no analog in GR due to the appearance of the dimensional parameter, $\epsilon$, which is the coefficient of  the higher order torsion tensor. This coefficient is not allowed to be zero, otherwise, the torsion scalar and the metric will be singular. The torsion scalar of this solution is constant, i.e. $T=\frac{-1}{6\epsilon}$.

The issue of singularities is discussed by calculating the scalars constructed from curvature and torsion. We have shown that all the scalars  will have a singularity at $\epsilon=0$ which ensure that this parameter must not equal to zero and ensure that our derived solution is a novel one. Moreover, we have shown that there is more singularities for the scalars constructed from torsion than those constructed from curvature. This addition singularity may be due to the divergence term which makes the torsion scalar differs from Ricci one. One more interesting property of solution (24) is that its Caushy horizon is not identical with horizon. This property shows the accumulation of the higher order torsion.

To investigate the physics of  Eq. (24) in a more  deep way we have calculated the energy and shows that it depends on  the parameter $\epsilon$. Equation (31) can not give the known form of energy due to the dimension parameter $\epsilon$ which demonstrates the effect of the higher torsion scalar. Furthermore, Eq. (31) shows that, in case of 4-dimension, if $\sqrt{c_4}=\frac{4\sqrt{2}}{27}$ and $c_1{}^2-c_2{}^2=1$ which gives $c_1=\sqrt{1+c_2{}^2}$ and  $E=\frac{c_3}{\epsilon}$. From this analysis we can relate $c_2$ to be the rotation parameter.

Then, we have discussed if solution (24) satisfies the first law of thermodynamics. We have shown that solution (24), which is an exact one to the non trivial case $f(T)=T+\epsilon T^2$, satisfied the first law of thermodynamics. This satisfaction comes from the fact that solution (24) gave a constant torsion. In general Miao et al. \cite{MLM} have shown that  for non constant  scalar torsion and when $f_{TT}\neq 0$   we get a violation of the first law of thermodynamics.

Finally, we want to make a comparison between our solution in which Caushy horizon of $g_{tt}=0$ and the horizons constructed from $g^{rr}=0$ are not the same. However, in orthodox GR the Caushy horizon of $g_{tt}=0$ and the horizons constructed from $g^{rr}=0$ are the same \cite{GPP}. This difference between the horizons of $g_{00}$ and $g^{rr}$ in higher order torsion at present time is not clear and its needs more study which will be done elsewhere.
\subsection*{Acknowledgments}
Authors would like to thank Referee for her/his useful comments which indeed improved the presentation of the paper and also we would like to thank Prof. A. Awad for fruitful discussion. This work is partially supported by the Egyptian Ministry of Scientific Research under project No. 24-2-12.


\end{document}